\documentclass[preprint]{article}
\usepackage{authblk}
\usepackage{graphicx}
\usepackage{dcolumn}
\usepackage{bm}
\usepackage{amsmath}
\usepackage{amssymb}
\usepackage{multirow}
\usepackage{caption}
\usepackage{subcaption}
\usepackage{epstopdf}
\usepackage{hyperref}

\title{\bf A novel approach to baryogenesis in $f(Q,L_{m})$ gravity and its cosmological implications}
\author[1]{Amit Samaddar\thanks{samaddaramit4@gmail.com}}
\author[2]{S. Surendra Singh\thanks{ssuren.mu@gmail.com}}
\affil[1,2]{Department of Mathematics, National Institute of Technology Manipur, Imphal-795004,India.}
\begin{document}

\maketitle 

\textbf{Abstract}: We present an examination of the $f(Q,L_{m})$ gravity model, in which the functional form $f(Q,L_{m})=\alpha Q^{n}+\beta L_{m}$ is postulated and discuss its potential impact on cosmological dynamics and the phenomenon of gravitational baryogenesis. Combining observational insights from Hubble, BAO and phantom datasets, we conduct a comprehensive analysis to constrain the model's parameters and determine the baryon-to-entropy ratio $\frac{\eta_{B}}{s}$, providing valuable insights into the model's performance and cosmological implications. In the context of baryogenesis and generalized gravitational baryogenesis, we show that setting $n=\frac{1}{2}$ results in a zero baryon-to-entropy ratio, which is physically implausible. Through a detailed examination of the dependence of $\frac{\eta_{B}}{s}$ on $n$ and $\beta$, we demonstrate that our model predicts a baryon-to-entropy ratio that is both positive and consistent with the observational upper limit of $9.42\times10^{-11}$ for $1.32965<n<1.39252$ and appropriate of $\beta$ and $n$ with $\alpha\simeq-1.95084\times10^{86}$. The excellent agreement between our model's predictions and the phantom dataset demonstrates the model's capacity to accurately describe the physics of baryogenesis and its ability to reproduce the observed features of the cosmological data, showcasing its potential as a reliable tool for understanding the evolution of the Universe.


\section{Introduction}\label{sec1}
\hspace{0.6cm} The theory of General Relativity (GR) has long reigned supreme as the definitive framework for understanding gravitational phenomena, demonstrating an uncanny ability to reconcile theoretical precision with observational accuracy across a stunning range of scales, from the minute to the majestic. Its enduring success has cemented its position as a cornerstone of modern astrophysics and cosmology, illuminating the intricacies of the Universe with unparalleled clarity. Through meticulous testing and validation, GR has consistently demonstrated its accuracy and reliability at various scales, yielding profound insights into the fabric and evolution of the Universe. Its predictive power has illuminated the dark corners of cosmic complexity, revealing the hidden patterns and processes that shape the Universe. A multitude of crucial cosmological observations, including Hubble's law, Big Bang nucleosynthesis and the Cosmic Microwave Background radiation, collectively attest to the applicability and accuracy of GR on cosmological scales, thereby solidifying its role as a fundamental pillar of modern cosmology \cite{E96}. Although GR has achieved remarkable triumphs, it encounters significant obstacles when confronted with contemporary astronomical observations, particularly at extended scales, revealing tensions between its predictions and the latest cosmic data. The convergence of high-precision data from Type Ia Supernovae (SNIa), large-scale structure surveys and galaxy cluster dynamics has brought to light a troubling discrepancy between GR's theoretical framework and the empirical reality of the Universe, prompting a re-evaluation of the theory's validity on cosmological scales \cite{Riess19,G13}. The discovery of the Universe's accelerating expansion, driven by an enigmatic component known as dark energy, has become a major puzzle in modern cosmology, requiring a concerted effort to unravel its secrets and reconcile it with our current understanding of the Universe \cite{S98}. The observed acceleration of the Universe's expansion confronts GR with a critical dilemma: either the theory requires refinement or augmentation, or an unknown entity, dark energy, is at play, necessitating a deeper comprehension of its nature and role in the Universe's evolution.

The observed predominance of matter over antimatter in the Universe, quantified as baryon asymmetry, represents a deeply intriguing enigma in the intersection of cosmology and particle physics, fueling ongoing research and theoretical exploration to unravel its underlying causes \cite{Rio99}. This refers to the overwhelming dominance of matter over antimatter, which appears inconsistent with the Standard Model of particle physics. In an effort to explain the observed baryon asymmetry, numerous theoretical models of baryogenesis have been proposed, exploring how this imbalance could have emerged during the Universe's early stages, including the radiation-dominated and matter-dominated eras, when the Universe was still in a state of rapid transformation \cite{M03}. Theoretical models of baryogenesis pivot on the concept of symmetry breaking, wherein the violation of charge (C) and charge-parity (CP) symmetry facilitates the asymmetric creation of matter and antimatter during particle interactions, ultimately giving rise to the Universe's preponderance of matter over antimatter \cite{SH06}. The Universe's asymmetry can only become pronounced once it exits the realm of thermal equilibrium and cools, as the equilibrium state would naturally give rise to a near-perfect balance between particles and antiparticles, masking the asymmetry \cite{S06}. The conditions required for baryogenesis were formalized by Sakharov \cite{AD67}. Fulfilling these conditions creates a window of opportunity for baryon asymmetry to emerge and evolve. New avenues of research have emerged, investigating the potential ties between baryogenesis and dark energy, with some theories proposing that the baryon asymmetry could be a consequence of the interactions between dark energy and the baryon sector \cite{Li02,Li03}. These theoretical models introduce a dynamic scalar field, commonly connected to dark energy, that interacts with the baryon current, giving rise to a spontaneous baryon asymmetry through a self-generated process.

In an innovative proposal, Davoudiasl et al. suggested a mechanism to generate baryon asymmetry without requiring the Universe to depart from thermal equilibrium, instead, leveraging a dynamical CP symmetry breaking via gravitational interaction \cite{H04}. The essential element of this model is the introduction of a CP-violating connection between the baryon current $J^{\mu}$ and the Ricci scalar's derivative $R$, which is encapsulated in the interaction term: $\frac{1}{M_{\star}^{2}}\int \sqrt{-g}J^{\mu}d^{4}x\partial_{\mu}(R)$, where $M_{\star}$ is a cutoff scale, typically associated with the reduced Planck mass, $M_{P}=\frac{1}{\sqrt{8\pi G}}\simeq2.4\times10^{18}$ $GeV$. By introducing this term, a gravitational contribution to baryogenesis is enabled, which becomes prominent at energy scales akin to quantum gravity, especially in the early Universe where such energies dominated. The relevance of this operator is further amplified in certain theories, where it arises naturally within the context of supergravity or higher-dimensional operators, courtesy of the K$\ddot{a}$hler potential, making it a key player in effective field theories that describe the low-energy behavior of quantum gravity.

GR has been incredibly successful, but its limitations in explaining certain cosmological observations have led to the development of modified gravity theories, which aim to build upon or depart from GR in order to provide a more accurate description of the Universe's large-scale behavior \cite{Clifton12}. Modified gravity theories provide a more comprehensive understanding of the Universe, addressing phenomena such as dark energy, dark matter and accelerating expansion, which are not fully captured by GR and shedding light on the underlying mechanisms driving the Universe's evolution. Numerous modifications to GR are obtained by modifying the Einstein-Hilbert action, which involves adding new terms that depend on various mathematical objects, such as curvature invariants, torsion, or non-metricity, thereby extending the original framework and enriching the theory's predictive power. The Riemann tensor forms the cornerstone of GR, the theory of gravity introduced by Albert Einstein in $1916$. One popular modification to GR is the $f(R)$ gravity theory, where the Einstein-Hilbert action is generalized by replacing the Ricci scalar $R$ with an arbitrary function $f(R)$ \cite{HA70}. Other geometric frameworks have surfaced as alternatives to GR, offering equivalent yet distinct formulations of gravitational dynamics. Within the realm of alternative geometric frameworks, Teleparallel Gravity (TEGR) \cite{CM61,K79} and Symmetric Teleparallel Gravity (STEGR) \cite{J18,B19} stand out as notable examples. TEGR replaces the curvature of spacetime with torsion, using a torsion scalar $T$ to describe gravitational interactions. Similarly, STEGR utilizes the concept of non-metricity, represented by the non-metricity scalar $Q$ as the key geometric quantity. By analogy with $f(R)$ gravity, the TEGR and STEGR frameworks can be generalized to yield $f(T)$ and $f(Q)$ gravity theories, which involve introducing arbitrary functions $f(T)$ and $f(Q)$ to replace the torsion scalar and non-metricity scalar, respectively \cite{Amit24,Singh23,N21}. Several additional techniques exist as well, including $f(R,T)$ \cite{T11}, $f(R, L_{m})$ \cite{T08,T10}, $f(T,\tau)$ \cite{T14}, $f(T,\phi)$ \cite{Amit2,L23}, $f(T,B)$ gravity \cite{Amit23} and $f(Q, T)$ \cite{S23}, to mention a number of them.

Moreover, scientists have explored further extensions of the theory by incorporating matter couplings, resulting in models such as $f(Q,L_{m})$ gravity \cite{Hazarika24}. In this framework, the matter sector is directly linked to the non-metricity scalar, introducing a new level of interaction between matter and geometry. Here, the Lagrangian $L_{m}$ embodies the Universe's matter content and its coupling with the non-metricity scalar $Q$ introduces a fresh interaction between the matter fields and the spacetime geometry, effectively merging the material and geometric facets of the Universe. This development creates new opportunities for investigating the behavior of gravitational and matter fields in a unified context. The incorporation of the arbitrary function $f(Q,L_{m})$ provides a theoretical framework with enhanced flexibility, enabling scientists to explore key cosmological puzzles, including the late-time acceleration of the Universe and the development of large-scale structures, with increased sophistication and accuracy. The model also provides a novel framework for addressing energy conditions and testing the theory's compatibility with observational data from large-scale structures and cosmic microwave background radiation. Studies on $f(Q,L_{m})$ gravity cover both analytical solutions and observational analysis, with \cite{Y24} providing a comprehensive discussion. Furthermore, \cite{M24} delves into the impact of bulk viscosity on this modified gravity theory. The phase space dynamics of BADE for $f(Q,L_{m})$ gravity are investigated in \cite{Ami24}.

Recently, researchers have explored the enigmatic concept of Baryogenesis within the framework of modified gravity theories. Various studies have investigated gravitational baryogenesis in different contexts, including $f(R)$ gravity \cite{G06}, Gauss-Bonnet gravity \cite{SD16}, $f(T)$ gravity \cite{V16}, $f(P)$ gravity \cite{SB21}, $f(R,T)$ gravity \cite{K18}, Ho$\check{r}$ava-Lifshitz \cite{N20}, $f(R,L_{m})$ gravity \cite{har15} and $f(Q,C)$ gravity \cite{M24}. In this paper, we delve into the cosmological implications of $f(Q,L_{m})$ gravity, focusing on its potential to provide a unified description of gravitational and matter interactions. By exploring specific forms of the function $f(Q,L_{m})$, we aim to investigate the role of this theory in addressing key issues such as the generation of baryon asymmetry, cosmic inflation and the dynamics of the early Universe. Through a detailed analysis of the field equations and observational constraints, we demonstrate that $f(Q,L_{m})$ gravity offers a rich and versatile framework for exploring new gravitational phenomena.

The structure of this paper is as follows: In Section \ref{sec2}, we begin by formulating the field equations for the $f(Q,L_{m})$ gravity model, establishing the core framework that describes the interaction between non-metricity and matter. Section \ref{sec3} explores the interpretation of the Hubble parameter within the context of $f(Q,L_{m})$ ravity, focusing on how this modified gravity theory affects cosmic expansion. Section \ref{sec4} addresses the observational constraints, where we analyze how the predictions of our model align with current cosmological observations. Section \ref{sec5} presents a detailed analysis of the baryogenesis mechanism in the context of $f(Q,L_{m})$ gravity, while Section \ref{sec6} extends this discussion to a generalized baryogenesis framework, highlighting its implications for cosmic evolution. Finally, Section \ref{sec7} provides a summary of the findings and concludes with a discussion of future research directions.
\section{Theoretical insights and field equations of $f(Q,L_{m})$ modified gravity}\label{sec2}
\hspace{0.5cm} This section delves into the mathematical structure of $f(Q,L_{m})$ gravity and extracts the associated field equations. Under this approach, the Riemann tensor provides a geometric description of gravity once a metric is specified. The Riemann tensor and its contractions play a crucial role in encapsulating the curvature of spacetime, providing insight into the gravitational interactions and how matter and energy influence the structure of spacetime.
\begin{equation}\label{1}
R^{a}_{b\mu\nu}=\partial_{\mu}Y^{a}_{\nu b}-\partial_{\nu}Y^{a}_{\mu b}+Y^{a}_{\mu\psi} Y^{\psi}_{\nu b}-Y^{a}_{\nu\psi} Y^{\psi}_{\mu b},
\end{equation}
We construct the Riemann tensor using an affine connection. In Weyl-Cartan geometry, an extension of Riemannian geometry, two important features emerge: torsion and non-metricity. These characteristics distinguish Weyl-Cartan geometry from its Riemannian counterpart. In this framework, the affine connection $Y^{a}_{\mu\nu}$ can be decomposed into three distinct components: the disformation tensor $L^{a}_{\mu\nu}$, the symmetric Levi-Civita connection $\Gamma^{a}_{\mu\nu}$ and contortion tensor $K^{a}_{\mu\nu}$. Therefore, the affine connection is expressed as follows:
\begin{equation}\label{2}
Y^{a}_{\mu\nu}=\Gamma^{a}_{\mu\nu}+L^{a}_{\mu\nu}+K^{a}_{\mu\nu}.
\end{equation}
The Levi-Civita connection $\Gamma^{a}_{\mu\nu}$ is a key element in differential geometry, representing a connection that is both torsion-free and compatible with the metric $g_{\mu\nu}$. It is entirely determined by the metric and its first derivatives, and it governs the curvature and parallel transport in general relativity. This connection encapsulates how spacetime curvature arises from gravitational interactions. Its explicit form is given by:
\begin{equation}\label{3}
\Gamma^{a}_{\mu\nu}=\frac{1}{2}g^{a\psi}(\partial_{\mu}g_{\psi\nu}+\partial_{\nu}g_{\psi\mu}-\partial_{\psi}g_{\mu\nu}),
\end{equation}
Measured geometrically, the departure from a torsion-free connection is given by the contortion tensor $K^{a}_{\mu\nu}$. Torsion's effect on spacetime is represented by this term, which is what differentiates between the Levi-Civita connection $\Gamma^{a}_{\mu\nu}$ and the affine connection $Y^{a}_{\mu\nu}$. The contortion tensor is expressed as:
\begin{equation}\label{4}
K^{a}_{\mu\nu}=\frac{1}{2}(T^{a}_{\mu\nu}+T_\mu{}^{a}{}_{\nu}+T_\nu{}^{a}{}_{\mu}),
\end{equation}
where $T^{a}_{\mu\nu}$ represents the torsion tensor, which encapsulates the twisting properties of spacetime. The torsion tensor leads to the torsion scalar $T$, which provides a measure of the extent to which the geometry deviates from being torsion-free. In gravitational theories involving torsion, the torsion scalar plays a critical role in governing the dynamics of the spacetime manifold.

The disformation tensor $L^{a}_{\mu\nu}$ quantifies the impact of non-metricity within a connection, capturing how the lengths of vectors change during parallel transport. It reflects the deviation from metric compatibility, indicating that the metric tensor is no longer preserved along transported paths. This tensor is fundamental in frameworks that extend beyond traditional Riemannian geometry, allowing for more general geometric structures of spacetime. It is defined as:
\begin{equation}\label{5}
L^{a}_{\mu\nu}=\frac{1}{2}(Q^{a}_{\mu\nu}-Q_\mu{}^{a}{}_{\nu}-Q_\nu{}^{a}{}_{\mu}),
\end{equation}
The non-metricity tensor, denoting the metric's deviation from covariantly constant, is $Q^{a}_{\mu\nu}$. The tensor of non-metricity is provided by:
\begin{equation}\label{6}
Q_{a\mu\nu}=\nabla_{a}g_{\mu\nu}=\partial_{a}g_{\mu\nu}-Y^{b}_{a\mu}g_{b\nu}-Y^{b}_{a\nu}g_{b\mu},
\end{equation}
This tensor reflects how non-metricity affects the geometry of spacetime by altering the metric under parallel transport. To incorporate a boundary term into the action of metric-affine gravitational theories, the introduction of the superpotential tensor $P^{a}_{\mu\nu}$, which is conjugate to the non-metricity, becomes necessary. This tensor plays a critical role in formulating boundary contributions that ensure the consistency of the action and uphold the variational principles. The superpotential is linked to the non-metricity tensor $Q^{a}_{\mu\nu}$, providing a means to account for the effects of non-metricity in a way that aligns with the geometry and physics of the theory. The superpotential tensor $P^{a}_{\mu\nu}$ is expressed as:
\begin{equation}\label{7}
P^{a}_{\mu\nu}=-\frac{1}{2}L^{a}_{\mu\nu}+\frac{1}{4}(Q^{a}-\widetilde{Q}^{a})g_{\mu\nu}-\frac{1}{4}\delta^{a}{}_{(\mu}Q_{\nu)},
\end{equation}
Here, $Q^{a}$=$Q^{a}{}_{\mu}{}^{\mu}$ and $\widetilde{Q}^{a}$=$Q_{\mu}{}^{a\mu}$ represent the non-metricity vectors. By contracting the non-metricity tensor with the superpotential tensor, we can derive the non-metricity scalar $Q$ as:
\begin{equation}\label{8}
Q=-Q_{\psi\mu\nu}P^{\psi\mu\nu}.
\end{equation}
The non-metricity scalar $Q$ measures the degree to which a manifold's geometry departs from the standard Riemannian framework. It indicates how the length or direction of an object varies during parallel transport, separate from any effects due to torsion. Specifically, $Q$ represents the inability of the metric to remain unchanged as an object moves through spacetime, highlighting the role that non-metricity plays in shaping the manifold’s overall geometry.

The $f(Q,L_{m})$ theory represents the gravitational action as follows:
\begin{equation}\label{9}
S=\int f(Q,L_{m})\sqrt{-g}d^{4}x,
\end{equation}
The terms involved in the equation (\ref{9}) are described as: $(a)$ The action $S$ is a mathematical representation of the gravitational system, incorporating the effects of both geometry and matter in spacetime and capturing their interconnected dynamics, $(b)$. The function $f(Q,L_{m})$ is a versatile mathematical expression that incorporates the non-metricity scalar $Q$ and the matter Lagrangian $L_{m}$, regulating the behavior of both gravity and matter in the Universe and presenting a holistic view of their interconnected dynamics, $(c)$ The symbol $\sqrt{-g}$  signifies the square root of the negative determinant of the metric tensor $g_{\mu\nu}$, ensuring that the action is invariant under general coordinate transformations and properly includes the volume element for curved spacetime.

By adjusting the action to account for changes in the metric tensor $g_{\mu\nu}$, the field equations take on the following precise form:
\begin{equation}\label{10}
\frac{2}{\sqrt{-g}}\nabla_{a}(f_{Q}\sqrt{-g}P^{a}_{\mu\nu})+f_{Q}(P_{\mu a\beta}Q_{\nu}{}^{a\beta}-2Q^{a\beta}{}_{\mu}P_{a\beta\nu})+\frac{1}{2}fg_{\mu\nu}=\frac{1}{2}f_{L_{m}}(g_{\mu\nu}L_{m}-T_{\mu\nu}),
\end{equation}
where $f_{Q}=\frac{\partial f}{\partial Q}$ and $f_{L_{m}}=\frac{\partial f}{\partial L_{m}}$. In this equation: $(a)$ The notation $\nabla_{a}$ stands for the covariant derivative, which captures the variation of a quantity under parallel transport in curved spacetime, reflecting the geometric properties of the spacetime manifold, $(b)$ The stress-energy tensor, denoted by $T_{\mu\nu}$, characterizes the arrangement of matter and energy within spacetime, providing a comprehensive description of their collective distribution. The stress-energy-momentum tensor is commonly expressed as:
\begin{equation}\label{11}
T_{\mu\nu}=-\frac{2}{\sqrt{-g}}\frac{\delta(\sqrt{-g}L_{m})}{\delta g^{\mu\nu}}=g_{\mu\nu}L_{m}-2\frac{\partial L_{m}}{\partial g^{\mu\nu}},
\end{equation}
The field equations emerge upon varying the action (\ref{9}) in response to changes in the connection, yielding:
\begin{equation}\label{12}
\nabla_{\mu}\nabla_{\nu}\bigg[4\sqrt{-g}f_{Q}P^{\mu\nu}_{a}+H^{\mu\nu}_{a}\bigg]=0,
\end{equation}
The hypermomentum density, represented by $H^{\mu\nu}$, incorporates the contributions of spin, dilation and shear from the matter fields, providing a generalized framework that surpasses the traditional stress-energy tensor. The mathematical expression for $H^{\mu\nu}$ is:
\begin{equation}\label{13}
H^{\mu\nu}=\sqrt{-g}f_{L_{m}}\frac{\delta L_{m}}{\delta Y^{a}_{\mu\nu}}.
\end{equation}
Upon taking the covariant derivative of the field equation (\ref{10}), we arrive at:
\begin{equation}\label{14}
D_{\mu}T^{\mu}_{\nu}=\frac{1}{f_{L_{m}}}\bigg(\frac{2}{\sqrt{-g}}\nabla_{a}\nabla_{\mu}H^{a\mu}_{\nu}+\nabla_{\mu}A^{\mu}_{\nu}-\nabla_{\mu}\bigg[\frac{1}{\sqrt{-g}}\nabla_{a}H^{a\mu}_{\nu}\bigg]\bigg    )=B_{\nu}\neq 0.
\end{equation}
The matter energy-momentum tensor $T_{\mu\nu}$ in $f(Q,L_{m})$ gravity experiences a non-standard evolution, characterized by the tensor $B_{\mu}$, which depends on the dynamic variables $Q$, $L_{m}$ and thermodynamic parameters. This departure from general relativity is a result of the interplay between geometry and matter, leading to a violation of the usual conservation laws. The equation $D_{\mu}T^{\mu}_{\nu}$=$B_{\nu}\neq 0$ encapsulates this non-conservation, underscoring the distinct dynamics in this theory.

To probe the cosmological implications of $f(Q,L_{m})$ gravity, we utilize the FLRW metric in Cartesian coordinates, which portrays a Universe with homogeneous and isotropic properties. The FLRW metric is given by:
\begin{equation}\label{15}
ds^{2}=-dt^{2}+a^{2}(t)(dx^{2}+dy^{2}+dz^{2}),
\end{equation}
In this context, the scale factor $a(t)$ a function of cosmic time, characterizes the Universe's expansion or contraction. It modifies the spatial coordinates, affecting the distances between objects over time. Within the FLRW framework, the non-metricity scalar $Q$ is expressed as $Q=6H^{2}$, where $H=\frac{\dot{a}}{a}$ represents the Hubble parameter. This scalar measures the deviation from metric preservation in the manifold's geometry and demonstrates how the expansion rate of the Universe influences non-metricity. The Hubble parameter $H$ quantifies the rate of change of the scale factor $a(t)$ over time, establishing a direct link between the Universe's geometric properties and its dynamic expansion behavior.

In the FLRW universe, filled with perfect fluid matter, the energy-momentum tensor $T_{\mu\nu}$ is crafted to capture the essence of this matter component. It is defined as:
\begin{equation}\label{16}
T_{\mu\nu}=(p+\rho)u_{\mu}u^{\nu}+pg_{\mu\nu},
\end{equation}
Here, $\rho$ embodies the energy density, a measure of the energy contained in a region of space. The pressure $p$ represents the force exerted per unit area, a consequence of the fluid's motion. The four-velocity $u_{\mu}$, encapsulates the relativistic velocity of the fluid, merging spatial and temporal aspects. This tensor $T_{\mu\nu}$ elegantly weaves together the properties of the perfect fluid, offering a unified portrayal of its energy and momentum distribution within the Universe. 

The modified Friedmann equations for $f(Q,L_{m})$ gravity govern the evolution of an FLRW Universe with perfect fluid matter, offering a brief understanding of the matter-geometry interplay. These equations are:
\begin{equation}\label{17}
3H^{2}=\frac{1}{4f_{Q}}\bigg[f-f_{L_{m}}(\rho+L_{m})\bigg],
 \end{equation}
 \begin{equation}\label{18}
\dot{H}+3H^{2}+\frac{\dot{f_{Q}}}{f_{Q}}H=\frac{1}{4f_{Q}}\bigg[f+f_{L_{m}}(p-L_{m})\bigg].
 \end{equation}
These equations serve as a foundation for exploring various cosmological and astrophysical scenarios, including large-scale structure evolution, cosmic acceleration and dark matter-dark energy interactions. By examining these equations, we can evaluate the $f(Q,L_{m})$ model's validity and compare it to observational evidence, potentially leading to groundbreaking insights into the nature of gravity. 
\section{Cosmological dynamics in the $f(Q,L_{m})=\alpha Q^{n}+\beta L_{m}$ model: motivation, implications and Hubble parameter analysis}\label{sec3}
\hspace{0.6cm} In this section, we assumed a specific model for $f(Q,L_{m})$ is given by:
\begin{equation}\label{19}
f(Q,L_{m})=\alpha Q^{n}+\beta L_{m},
\end{equation}
where $\alpha$ and $\beta$ are constants. This model is driven by the ambition to investigate generalized theories of gravity that integrate both geometric and matter-related effects, enabling a more complex and dynamic relationship between the fabric of spacetime and the Universe's material composition. The power-law form $Q^{n}$ provides a versatile framework for investigating departures from GR and exploring possible corrective mechanisms that may arise during different eras of the Universe's development. By incorporating the matter Lagrangian $L_{m}$, the Universe's matter content is explicitly integrated into the gravitational dynamics, enabling a direct interaction between matter and geometry. This coupling gives rise to novel phenomenological consequences, including altered energy conservation laws and non-standard matter interactions that deviate from the predictions of GR. By adjusting the parameter $n$ to certain values, the model can mimic the accelerated expansion of the Universe, eliminating the requirement for a cosmological constant or dark energy and offering a fresh perspective on this long-standing cosmological puzzle.

In the case where $n=1$, $\alpha=1$ and $\beta=0$, the model collapses to GR, as $Q$ becomes identical to the Ricci scalar $R$ and the matter Lagrangian takes on its conventional form. This approach is in line with other modified gravity theories, such as $f(T)$ gravity, which considers a power-law form $f(T)=\alpha T^{n}$ \cite{M12}, $f(Q,T)$ gravity, which assumes a model $f(Q,T)=\alpha Q^{n+1}+\beta T$ \cite{Xu19}, $f(T,\tau)$ gravity, which proposes a model $f(T,\tau)=\alpha T^{n}\tau+\Lambda$ \cite{T24} and $f(R,L_{m})$ gravity, which adopts a form $f(R,L_{m})=\frac{R}{2}+L_{m}^{n}+\beta$ \cite{L22}. \\

$\bullet$ \textbf{Implications of the hubble parameter:} Upon inserting the model $f(Q,L_{m})=\alpha Q^{n}+\beta L_{m}$ into the field equations (\ref{17}) and (\ref{18}), we obtained the Hubble parameter $H(t)$ as:
\begin{equation}\label{20}
H(t)=\frac{2n}{3(t-t_{0})},
\end{equation}
This form suggests that the Universe's expansion rate is governed by the parameter $n$, which regulates the extent of non-linearity in the model, thereby influencing the cosmic evolution. The inverse dependence on $t$ suggests that as time progresses, the Hubble parameter decreases, indicating a decelerating Universe for $n>1$, or a specific rate of expansion for different values of $n$. This behavior aligns with the dynamics of the early Universe and offers a window into various phases of cosmic evolution, encompassing inflationary epochs, late-time deceleration, or acceleration, depending on the value of $n$, which serves as a cosmic evolution phase switch.

By applying the formula for the time derivative of the Hubble parameter, $\dot{H}=-(1+z)H(z)\frac{dH}{dz}$, we can rewrite the Hubble parameter in terms of redshift as:
\begin{equation}\label{21}
H(z)=H_{0}(1+z)^{\frac{3}{2n}},
\end{equation}
This equation provides a vital connection between the Hubble parameter $H_{0}$, which describes the Universe's current expansion rate and the redshift $z$, allowing for a direct and meaningful comparison with observational data, and thereby testing the validity of cosmological theories. The value of $n$ dictates the power-law dependence on $z$, resulting in a multiplicity of cosmological outcomes, as different $n$ values give rise to distinct expansion histories, enriching our understanding of the Universe's evolution. Specifically, when $n=1$, the equation simplifies to the classic $H(z)\propto (1+z)^{\frac{3}{2}}$ relationship, recovering the standard GR cosmological model, which characterizes a Universe dominated by matter with a predictable expansion evolution. In contrast, when $n\neq1$, the model's predictions for the Hubble parameter's evolution stray from the standard model's expectations, resulting in distinct forecasts for cosmic expansion at high redshifts. This discrepancy provides a chance to rigorously test the model against observational data from supernovae, cosmic microwave background radiation and baryon acoustic oscillations, enabling a precise assessment of its accuracy.
\section{Observational constraints on the $f(Q,L_{m})$ model}\label{sec4}
\hspace{0.6cm} In this section, we leverage multiple observational datasets to impose constraints on the model's parameters, enabling a more precise calibration of the model and a deeper insight into its accuracy. We employ advanced statistical tools, specifically Markov Chain Monte Carlo (MCMC) techniques, as implemented in the emcee Python package \cite{D13}, to precisely constrain key parameters like $H_{0}$ and $n$ using established Bayesian inference methods, enabling a robust and probabilistic estimation of their values. In order to fine-tune the best-fit values for the model parameters, we also employ the probability function, enabling an accurate and reliable estimation of their values shown below:
\begin{equation}\label{22}
\mathcal{L}\propto e^{\frac{-\chi^{2}}{2}},
\end{equation}
where $\chi^{2}$ stands for the chi-squared function, a mathematical entity used to calculate the squared difference between observed and expected values, providing a metric for assessing the model's fit to the data.Our study centers on three pivotal datasets: $H(z)$ measurements of the Hubble parameter, BAO and Phantom. Moreover, we constrain the parameters with the following prior intervals: $60.0 < H_{0} < 80.0$ and $0 < n < 5$, to identify the regime of accelerated expansion and elucidate its characteristics. In the MCMC framework, multiple chains are initiated from random locations within the parameter space and then navigate the space by iteratively sampling from the likelihood function, enabling a thorough exploration of the parameter space and the underlying probability distribution. The $\chi^{2}$ function used to analyze various datasets is defined as follows:
\subsection{Hubble dataset}\label{sec4.1}
\begin{figure}[h!]
\centering
  \includegraphics[scale=0.4]{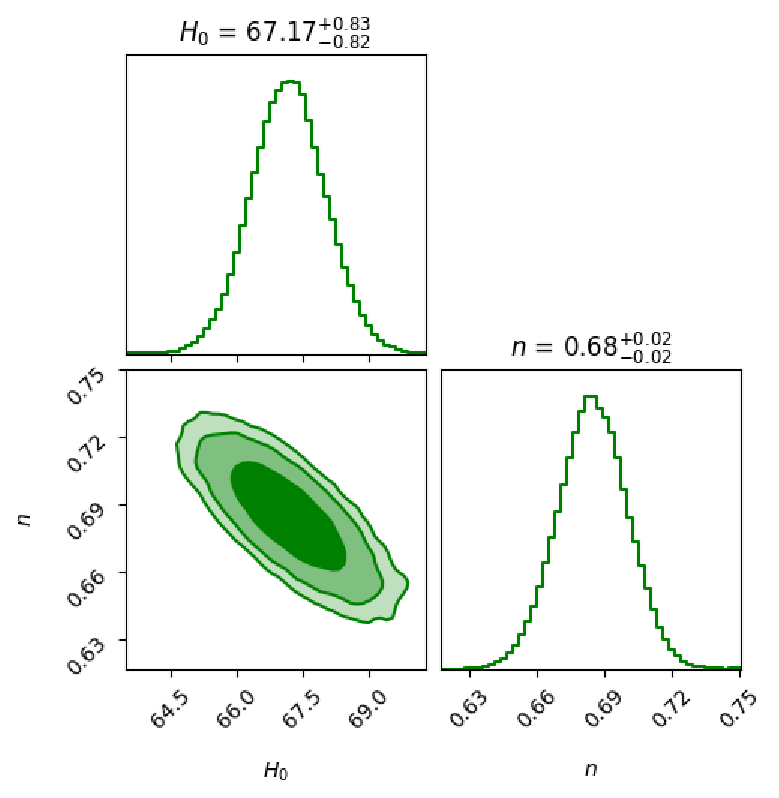}
  \caption{The contour plots display the confidence regions for the model parameters at $1-\sigma$ and $2-\sigma$ levels, derived from the Hubble dataset.}
  \label{fig:f1}
\end{figure}

\begin{figure}[h!]
\centering
  \includegraphics[scale=0.4]{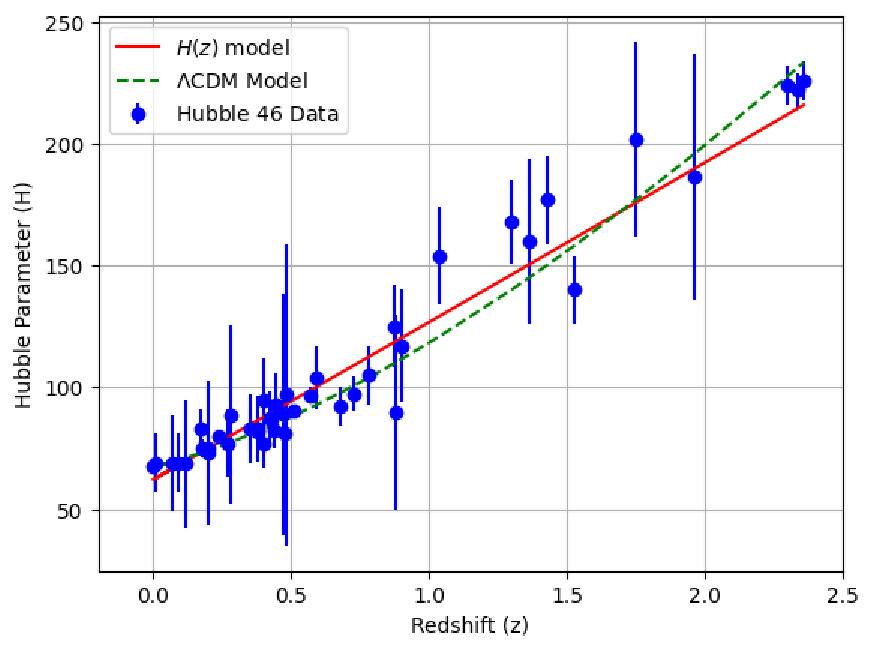}
  \caption{The plot features $46$ blue data points, each representing an observational Hubble parameter measurement, along with its corresponding error bar, which graphically illustrates the uncertainty in the data.}
  \label{fig:f2}
\end{figure}

\hspace{0.6cm} Utilizing the Cosmic Chronometers (CC) method, researchers determine the Hubble rate by studying galaxies that are ancient, passively evolving and separated by minimal redshift intervals, employing the differential age method \cite{D10,M12} to extract valuable insights. This approach capitalizes on the fact that the age disparities between these galaxies are directly tied to the Hubble parameter at distinct redshifts, enabling a model-independent and unbiased reconstruction of the Universe's expansion history. The technique relies on the following formula of the Hubble rate
\begin{equation}\label{23}
H(z)=-\frac{1}{(1+z)}\frac{dz}{dt},
\end{equation}
Spectroscopic surveys provide the change in redshift $(dz)$, which, when combined with the measurement of the change in time $(dt)$, yields the Hubble parameter's value independent of any specific model, offering a valuable insight into the cosmic expansion. We drew upon an exhaustive set of $46$ Hubble parameter measurements, covering a redshift range of $0$ to $2.36$, to calibrate our cosmological model. A comprehensive summary of these data points and their associated references is provided in Table 1. The chi-square is given by:
\begin{equation}\label{24}
\chi^{2}_{H}=\sum_{i=1}^{46}\frac{[H_{th}(p,z_{i})-H_{obs}(z_{i})]^{2}}{[\sigma_{H}(z_{i})]^{2}}.
\end{equation}
The equation features the theoretical Hubble parameter value, $H_{th}(p,z_{i})$ , which is the model's prediction for the Hubble parameter at a given redshift $z_{i}$, with $p$ being the set of model parameters being refined. The observed value of the Hubble parameter at the same redshift $z_{i}$ is $H_{obs}(z_{i})$ and the corresponding standard error, $\sigma_{H}(z_{i})$ provides a measure of the uncertainty in that observation. The Hubble data, along with their uncertainties, are graphically presented in Figure \ref{fig:f2}, while Figure \ref{fig:f1} displays contour plots for the model parameters, highlighting the confidence regions at $1-\sigma$ and $2-\sigma$ levels. This analysis allows us to set robust limits on the cosmological parameters and assess the model's ability to accurately describe the Universe's expansion evolution.
\begin{table}[h!]
\centering
\caption{$46$ datasets of $H(z)$}
\begin{tabular}{||p{1.3cm}|p{1.3cm}|p{1.3cm}|p{0.6cm}||p{1.3cm}|p{1.6cm}|p{0.6cm}|p{1cm}||}
\hline\hline
 $\hspace{0.4cm}z_{i}$ & $\hspace{0.2cm} H_{obs}$ & $\hspace{0.3cm}\sigma_{H}$ & Ref. & $\hspace{0.4cm}z_{i}$ & $\hspace{0.2cm}H_{obs}$ & $\hspace{0.2cm}\sigma_{H}$ & Ref. \\
\hline\hline
$\hspace{0.4cm}0$ & $\hspace{0.2cm}67.77$ & $\hspace{0.2cm}1.30$ & \cite{Nichol19} & $\hspace{0.2cm}0.4783$ & $\hspace{0.2cm}80.9$ & $\hspace{0.2cm}9$ & \cite{Moresco16}\\
\hline
$\hspace{0.2cm}0.07$ & $\hspace{0.4cm}69$ & $\hspace{0.2cm} 19.6$ & \cite{Zhang14} & $\hspace{0.2cm}0.48$ & $\hspace{0.2cm}97$ & $\hspace{0.2cm} 60$ & \cite{Stern10} \\
\hline
$\hspace{0.2cm}0.09$ & $\hspace{0.4cm}69$ & $\hspace{0.3cm} 12$ & \cite{Simon05} & $\hspace{0.2cm}0.51$ & $\hspace{0.2cm}90.4$ & $\hspace{0.2cm} 1.9$ & \cite{Alam16} \\
\hline
$\hspace{0.2cm}0.01$ & $\hspace{0.4cm}69$ & $\hspace{0.3cm} 12$ & \cite{Stern10} & $\hspace{0.2cm}0.57$ & $\hspace{0.2cm}97$ & $\hspace{0.2cm} 3.4$ & \cite{Cimatti12} \\
\hline
$\hspace{0.2cm}0.12$ & $\hspace{0.4cm}68.6$ & $\hspace{0.2cm} 26.2$ & \cite{Zhang14} & $\hspace{0.2cm}0.59$ & $\hspace{0.2cm}104$ & $\hspace{0.2cm} 13$ & \cite{Alam16}\\
\hline
$\hspace{0.2cm}0.17$ & $\hspace{0.4cm}83$ & $\hspace{0.4cm} 8$ & \cite{Stern10} & $\hspace{0.2cm}0.60$ & $\hspace{0.2cm}87.6$ & $\hspace{0.2cm} 6.1$ & \cite{Blake12}\\
\hline
$\hspace{0.2cm}0.179$ & $\hspace{0.4cm}75$ & $\hspace{0.4cm} 4$ & \cite{Cimatti12} & $\hspace{0.2cm}0.61$ & $\hspace{0.2cm}97.3$ & $\hspace{0.2cm} 2.1$ & \cite{Alam16} \\
\hline
$\hspace{0.2cm}0.1993$ & $\hspace{0.4cm}75$ & $\hspace{0.4cm} 5$ & \cite{Cimatti12} & $\hspace{0.2cm}0.68$ & $\hspace{0.2cm}92$ & $\hspace{0.2cm} 8$ & \cite{Cimatti12} \\
\hline
$\hspace{0.2cm}0.20$ & $\hspace{0.4cm}72.9$ & $\hspace{0.2cm} 29.6$ & \cite{Zhang14} & $\hspace{0.2cm}0.73$ & $\hspace{0.2cm}97.3$ & $\hspace{0.2cm} 7$ & \cite{Blake12} \\
\hline
$\hspace{0.2cm}0.24$ & $\hspace{0.4cm}79.7$ & $\hspace{0.2cm} 2.7$ & \cite{Cabre09} & $\hspace{0.2cm}0.781$ & $\hspace{0.2cm}105$ & $\hspace{0.2cm} 12$ & \cite{Cimatti12} \\
\hline
$\hspace{0.2cm}0.27$ & $\hspace{0.4cm}77$ & $\hspace{0.2cm} 14$ & \cite{Stern10} & $\hspace{0.2cm}0.875$ & $\hspace{0.2cm}125$ & $\hspace{0.2cm} 17$ & \cite{Cimatti12} \\
\hline
$\hspace{0.2cm}0.28$ & $\hspace{0.4cm}88.8$ & $\hspace{0.2cm} 36.6$ & \cite{Zhang14} & $\hspace{0.2cm}0.881$ & $\hspace{0.2cm}90$ & $\hspace{0.2cm} 40$ & \cite{Stern10} \\
\hline
$\hspace{0.2cm}0.35$ & $\hspace{0.4cm}82.7$ & $\hspace{0.2cm} 8.4$ & \cite{Wang13} & $\hspace{0.2cm}0.9$ & $\hspace{0.2cm}117$ & $\hspace{0.2cm} 23$ & \cite{Stern10} \\
\hline
$\hspace{0.2cm}0.352$ & $\hspace{0.4cm}83$ & $\hspace{0.2cm} 14$ & \cite{Cimatti12} & $\hspace{0.2cm}1.037$ & $\hspace{0.2cm}154$ & $\hspace{0.2cm} 20$ & \cite{Cimatti12} \\
\hline
$\hspace{0.2cm}0.38$ & $\hspace{0.4cm}81.5$ & $\hspace{0.2cm} 1.9$ & \cite{Alam16} & $\hspace{0.2cm}1.3$ & $\hspace{0.2cm}168$ & $\hspace{0.2cm} 17$ & \cite{Stern10} \\
\hline
$\hspace{0.2cm}0.3802$ & $\hspace{0.4cm}88.8$ & $\hspace{0.2cm} 36.6$ & \cite{Moresco16} & $\hspace{0.2cm}1.363$ & $\hspace{0.2cm}160$ & $\hspace{0.2cm} 33.6$ & \cite{Michele16} \\
\hline
$\hspace{0.2cm}0.4$ & $\hspace{0.4cm}95$ & $\hspace{0.2cm} 17$ & \cite{Simon05} & $\hspace{0.2cm}1.43$ & $\hspace{0.2cm}177$ & $\hspace{0.2cm} 18$ & \cite{Stern10} \\
\hline
$\hspace{0.2cm}0.4004$ & $\hspace{0.4cm}77$ & $\hspace{0.2cm} 10.2$ & \cite{Moresco16} & $\hspace{0.2cm}1.53$ & $\hspace{0.2cm}140$ & $\hspace{0.2cm} 14$ & \cite{Stern10} \\
\hline
$\hspace{0.2cm}0.4247$ & $\hspace{0.4cm}87.1$ & $\hspace{0.2cm} 11.2$ & \cite{Moresco16} & $\hspace{0.2cm}1.75$ & $\hspace{0.2cm}202$ & $\hspace{0.2cm} 40$ & \cite{Stern10} \\
\hline
$\hspace{0.2cm}0.43$ & $\hspace{0.4cm}86.5$ & $\hspace{0.2cm} 3.7$ & \cite{Cabre09} & $\hspace{0.2cm}1.965$ & $\hspace{0.2cm}186.5$ & $\hspace{0.2cm} 50.4$ & \cite{Michele16} \\
\hline
$\hspace{0.2cm}0.44$ & $\hspace{0.4cm}82.6$ & $\hspace{0.2cm} 7.8$ & \cite{Blake12} & $\hspace{0.2cm}2.3$ & $\hspace{0.2cm}224$ & $\hspace{0.2cm} 8$ & \cite{Rich13} \\
\hline
$\hspace{0.2cm}0.44497$ & $\hspace{0.4cm}92.8$ & $\hspace{0.2cm} 12.9$ & \cite{Moresco16} & $\hspace{0.2cm}2.34$ & $\hspace{0.2cm}222$ & $\hspace{0.2cm} 7$ & \cite{Delubac15} \\
\hline
$\hspace{0.2cm}0.47$ & $\hspace{0.4cm}89$ & $\hspace{0.2cm} 49.6$ & \cite{Cress17} & $\hspace{0.2cm}2.36$ & $\hspace{0.2cm}226$ & $\hspace{0.2cm} 8$ & \cite{Andreu14} \\
\hline
\end{tabular}
\label{Tab:T1}
\end{table}
\subsection{BAO dataset}\label{4.2}
\hspace{0.6cm} Our model is fine-tuned using BAO, which exploits the regular fluctuations in the density of visible baryonic matter, providing a precise tool to probe the Universe's expansion history. This method enables us to constrain our model and uncover the secrets of the cosmos. To constrain our model using BAO, we employ the acoustic scale $l_{A}$ as:
\begin{equation}\label{25}
l_{A}=\frac{\pi d_{A}(z_{d})}{r_{s}(z_{\star})},
\end{equation}
where $d_{A}(z)=c\int_{0}^{z}\frac{dz'}{H(z')}$ denotes the distance between objects in the Universe, measured in comoving coordinates. The sound horizon, $r_{s}$ is a cosmological distance scale that marks the furthest distance that sound waves could have traveled in the Universe determined by
\begin{equation}\label{26}
r_{s}=\int_{z_{d}}^{\infty}\frac{c_{s}(z')}{H(z)},
\end{equation}
at the drag epoch $z_{d}$ and $c_{s}(z')$ represents the sound speed. The following expression yields the dilation scale $D_{V}(z)$:
\begin{equation}\label{27}
D_{V}(z)=\bigg(\frac{d_{A}^{2}(z)cz}{H(z)}\bigg)^{\frac{1}{3}},
\end{equation}
The dilation scale helps characterize the large-scale structure and derive cosmic facts from the BAO signal because it combines both radial and transverse distances. To assess the BAO fit, the Chi-square function is constructed over the redshift interval $0.24<z<2.36$ as follows:
\begin{equation}\label{28}
\chi^{2}_{BAO}=\sum_{i=1}^{15}\bigg[\frac{D_{obs}-D_{th}(z_{i})}{\Delta D_{i}}\bigg]^{2}.
\end{equation}
In this expression, $D_{th}$ refers to the theoretical distance measure predicted by the model at redshift $z_{i}$, while $D_{obs}$ is the observed distance measure. The term $\Delta D_{i}$ represents the uncertainty in the observed distance measure at each redshift. The analysis incorporates $15$ BAO data points, listed in Table \ref{Tab2}, along with their respective observational sources. This Chi-square approach allows for the comparison of theoretical predictions with observational data to constrain the model parameters.
\begin{table}
\centering
\caption{$15$ datasets of BAO and other method}
\begin{tabular}{||p{1.3cm}|p{1.3cm}|p{1.3cm}|p{0.6cm}||}
\hline\hline
 $\hspace{0.5cm}z_{i}$ & $\hspace{0.3cm} H_{obs}$ & $\hspace{0.3cm}\sigma_{H}$ & Ref.\\
\hline\hline
$\hspace{0.4cm}0.30$ & $\hspace{0.2cm}81.77$ & $\hspace{0.2cm}6.22$ & \cite{Oka14}\\
\hline
$\hspace{0.4cm}0.31$ & $\hspace{0.2cm}78.18$ & $\hspace{0.2cm} 4.74$ & \cite{E09}\\
\hline
$\hspace{0.4cm}0.34$ & $\hspace{0.2cm}83.8$ & $\hspace{0.2cm} 3.66$ & \cite{E09}\\
\hline
$\hspace{0.4cm}0.36$ & $\hspace{0.2cm}79.94$ & $\hspace{0.2cm} 3.38$ & \cite{Y17} \\
\hline
$\hspace{0.4cm}0.40$ & $\hspace{0.2cm}82.04$ & $\hspace{0.2cm} 2.03$ & \cite{Y17}\\
\hline
$\hspace{0.4cm}0.43$ & $\hspace{0.2cm}86.45$ & $\hspace{0.2cm} 3.97$ & \cite{E09}\\
\hline
$\hspace{0.4cm}0.44$ & $\hspace{0.2cm}84.81$ & $\hspace{0.2cm} 1.83$ & \cite{Blake12}\\
\hline
$\hspace{0.4cm}0.48$ & $\hspace{0.2cm}87.79$ & $\hspace{0.2cm} 2.03$ & \cite{Blake12} \\
\hline
$\hspace{0.4cm}0.52$ & $\hspace{0.2cm}94.35$ & $\hspace{0.2cm} 2.64$ & \cite{Blake12} \\
\hline
$\hspace{0.4cm}0.56$ & $\hspace{0.2cm}93.34$ & $\hspace{0.2cm} 2.3$ & \cite{Blake12} \\
\hline
$\hspace{0.4cm}0.57$ & $\hspace{0.2cm}87.6$ & $\hspace{0.2cm} 7.8$ & \cite{C13} \\
\hline
$\hspace{0.4cm}0.59$ & $\hspace{0.2cm}98.48$ & $\hspace{0.2cm} 3.18$ & \cite{L14} \\
\hline
$\hspace{0.4cm}0.61$ & $\hspace{0.2cm}97.3$ & $\hspace{0.2cm} 2.1$ & \cite{The17} \\
\hline
$\hspace{0.4cm}0.64$ & $\hspace{0.2cm}98.82$ & $\hspace{0.2cm} 2.98$ & \cite{Busca13} \\
\hline
$\hspace{0.4cm}2.33$ & $\hspace{0.2cm}224$ & $\hspace{0.4cm} 8$ & \cite{The17} \\
\hline\hline
\end{tabular}
\label{Tab2}
\end{table}
\subsection{MCMC-based joint inference of Hubble and BAO Data}\label{sec4.3}
\begin{figure}[h!]
\centering
  \includegraphics[scale=0.4]{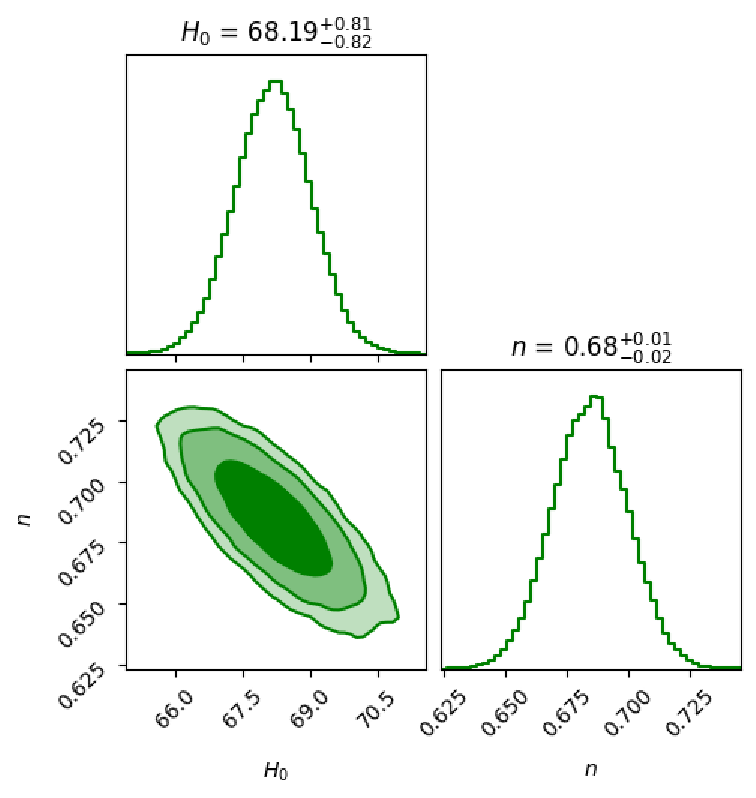}
  \caption{$1-\sigma$ and $2-\sigma$ confidence regions for model parameters, based on Hubble+BAO dataset.}
  \label{fig:f3}
\end{figure}

\begin{figure}[h!]
\centering
  \includegraphics[scale=0.4]{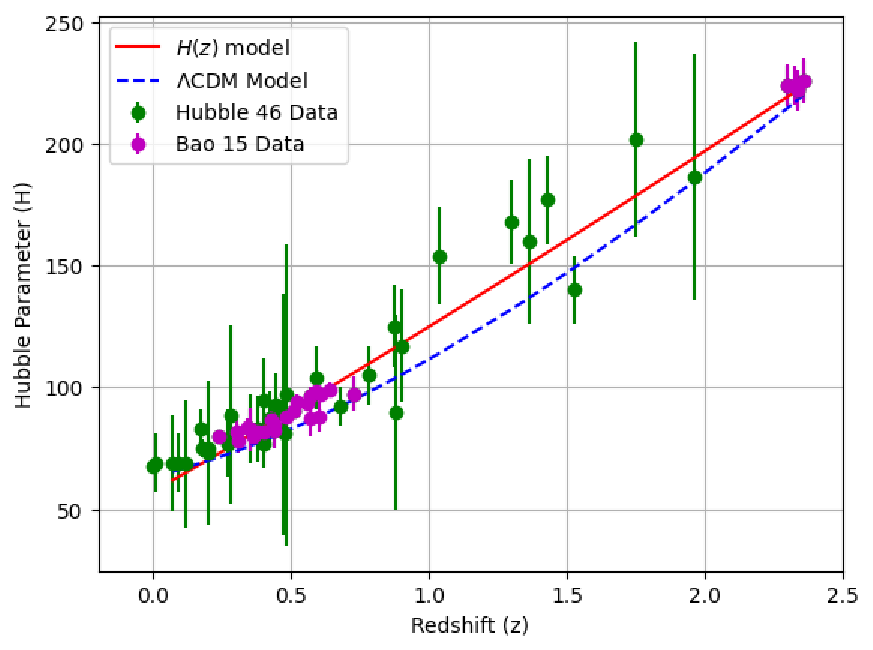}
  \caption{The plot illustrates $H$ vs $z$ for our model, with green and pink dots representing Hubble and BAO data points, respectively, with associated error bars.}
  \label{fig:f4}
\end{figure}

\hspace{0.6cm} Utilizing the capabilities of MCMC analysis, we jointly analyze the Hubble and BAO datasets to improve parameter estimates in the $f(Q,L_{m})$ gravity model. By merging these datasets, we capitalize on their synergies to achieve a more precise characterization of model parameters. The total chi-square statistic is computed by aggregating the individual contributions from Hubble ($\chi^{2}_{H}$) and BAO ($\chi^{2}_{BAO}$) datasets, yielding:
\begin{equation}\label{29}
 \chi^{2}=\chi^{2}_{H}+\chi^{2}_{BAO}.
 \end{equation}
 The combined dataset's confidence levels are visualized in Figure \ref{fig:f3}, with contour lines representing the $1-\sigma$ and $2-\sigma$ ranges. Also, in Figure \ref{fig:f4}, an error bar plot of $H$ versus $z$ is shown, with green and pink markers denoting Hubble and BAO data points, respectively. By analyzing the data jointly, we gain a more comprehensive understanding of the model's performance, leading to more robust constraints on cosmological parameters.
\subsection{Phantom dataset}\label{sec4.4}
\hspace{0.6cm} SNe Ia, with its standard glowing character, is crucial for restricting the dark energy sector as exact distance indicators. Our research is based on $1048$ data points from the Pantheon collection, which spans $0.01$ to $2.26$ in redshift \cite{Ak20,Dm18}. The chi-squared statistic for this dataset is given by
\begin{equation}\label{30}
\chi^{2}_{Phantom}=\sum_{i=1}^{1048}\bigg[\frac{\mu_{obs}(z_{i})-\mu_{th}(z_{i})}{\sigma_{\mu} (z_{i})}\bigg]^{2}.
\end{equation}
where $\mu_{obs}(z_{i})$ is the data point, $\mu_{th}(z_{i})$ is the model's prediction and $\sigma_{\mu} (z_{i})$ is the uncertainty in the data point. Moreover, we introduce $\mu$, which represents the contrast between the observed apparent magnitude $(m_{B})$ and the absolute magnitude $(M_{B})$ at a given redshift, $i.e.$, $\mu=m_{B}-M_{B}$. The model's theoretical distance modulus is given by $\mu(z)$ as:
\begin{equation}\label{31}
\mu(z)=5\log_{10}\bigg(\frac{D_{L}}{H_{0}Mpc}\bigg)+25.
\end{equation}
where $D_{L}(z)$ is the distance light could travel in a flat universe, expressed as
\begin{equation}\label{32}
D_{L}(z)=c(1+z)\int_{0}^{z}\frac{dz'}{H(z')}.
\end{equation}

\begin{figure}[h!]
\centering
  \includegraphics[scale=0.4]{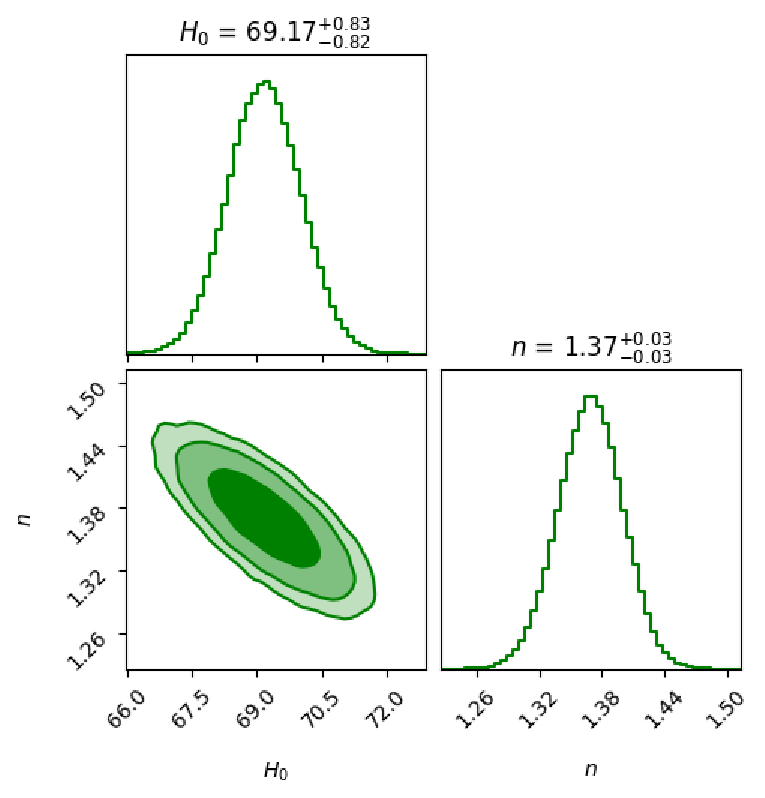}
  \caption{$1-\sigma$ and $2-\sigma$ confidence regions for model parameters, based on Hubble+Phantom dataset.}
  \label{fig:f5}
\end{figure}

We expand our analysis by combining the Hubble and Phantom datasets and applying the MCMC method. This joint analysis refines our parameter estimates for the $f(Q,L_{m})$ gravity model by incorporating additional data, providing more comprehensive observational constraints. The resulting chi-square function is the sum of the individual contributions from the Hubble and Phantom datasets:
\begin{equation}\label{33}
 \chi^{2}=\chi^{2}_{H}+\chi^{2}_{Phantom}.
\end{equation}
Figure \ref{fig:f5} demonstrates the contour plot establishing the $1-\sigma$ and $2-\sigma$ regions for the merged Hubble and Phantom datasets.
\begin{table}[h!]
\centering
\caption{Optimal parameter values that provide the best fit }
\begin{tabular}{||p{4.1cm}|p{2.0cm}|p{2.0cm}||}
\hline\hline
 Parameters & \hspace{0.6cm}$H_{0}$ & \hspace{0.9cm}$n$ \\
\hline\hline
$\hspace{0.3cm}OHD46$ & $67.17^{+0.83}_{-0.82}$ & $0.68^{+0.02}_{-0.02}$\\[1pt]
\hline
$OHD46+BAO15$ & $68.19^{+0.81}_{-0.82}$ & $0.68^{+0.01}_{-0.02}$\\ [1pt]
\hline
$OHD46+Phantom1048$ & $69.22^{+0.83}_{-0.83}$ & $1.46^{+0.03}_{-0.03}$\\ [1pt]
\hline\hline
\end{tabular}
\label{Tab:T3}
\end{table}

The estimated values of the model parameters are:\\
$\bullet$ Hubble dataset: $H_{0}= 67.17^{+0.83}_{-0.82}$ $km/s/Mpc$ and $n=0.68^{+0.02}_{-0.02}$,\\
$\bullet$ Joint Hubble and BAO Dataset: $H_{0}= 68.19^{+0.81}_{-0.82}$ $km/s/Mpc$ and $n=0.68^{+0.01}_{-0.02}$,\\
$\bullet$ Joint Hubble and Phantom Dataset: $H_{0}= 69.22^{+0.83}_{-0.83}$ $km/s/Mpc$ and $n=1.46^{+0.03}_{-0.03}$.\\

The $H_{0}$ values are consistent with the $\Lambda$CDM model, which predicts $H_{0}\approx 65-75,  \text{km/s/Mpc}$. The $H_{0}$ value derived from both the Hubble dataset and the combined Hubble-BAO dataset aligns closely with the current standard value obtained from the Planck mission ($\approx 67.4 , \text{km/s/Mpc}$) \cite{Planck20}, reinforcing the model’s compatibility with standard cosmological observations. On the other hand, the slightly elevated $H_{0}$ value from the joint Hubble-Phantom dataset suggests it bridges the gap between the standard $\Lambda$CDM result and other measurements reporting a higher Hubble constant, such as \cite{Wl19}. This intermediate value could reflect differences in how phantom energy interacts with cosmic expansion, hinting at new insights into late-time acceleration. The parameter $n$ remains consistent within the joint Hubble and BAO dataset and the Hubble dataset, indicating that the model predicts similar values for the cosmic evolution parameter $n$ regardless of slight variations in the data. However, the combined Hubble and Phantom dataset suggests a significantly larger value for $n$, implying potential deviations or corrections required for models involving phantom fields.
\section{Baryogenesis in $f(Q,L_{m})$ gravity}\label{sec5}
\hspace{0.6cm} This section examines the possibility of gravitational baryogenesis in $f(Q,L_{m})$ gravity, focusing on how this alternative gravity theory can provide insights into the matter-antimatter asymmetry observed in the Universe. Baryogenesis is the scientific term for the mysterious process that created an imbalance between matter and antimatter in the Universe, resulting in the existence of stars, planets and life as we know it. The baryon-to-entropy ratio is a fundamental measure of the Universe's matter-antimatter asymmetry, representing the number of baryons relative to the total entropy as follows:
\begin{equation}\label{34}
\eta_{B}=\frac{n_{B}-\tilde{n}_{B}}{s},
\end{equation}
In this case, $n_{B}$ denotes the baryon quantity, $\tilde{n}_{B}$ represents the anti-baryon quantity and $s$ denotes the Universe's entropy. The observational evidence from Big Bang Nucleosynthesis (BBN) and Cosmic Microwave Background (CMB) radiation has set a tight constraint on the baryon-to-entropy ratio, limiting its value to a narrow range around $9\times 10^{-11}$, which provides valuable insights into the Universe's matter-antimatter asymmetry \cite{C03,S01}. 

The formation of the matter-antimatter asymmetry in the circumstances of baryogenesis is significantly influenced by the Sakharov requirements \cite{A67}. The three crucial prerequisites for baryogenesis to occur are: $(1)$ baryon number must be violated, $(2)$ both charge (C) and charge-parity (CP) symmetries must be broken and $(3)$ the process must occur in a state of non-thermal equilibrium, deviating from the equilibrium state of the Universe. The expanding Universe leads to a decrease in temperature $(T)$ and once it crosses a threshold temperature $(T_{D})$, the baryogenesis processes cease to operate, leaving a residual baryon-to-entropy ratio as a lasting imprint on the Universe. Theoretical models of gravitational baryogenesis provide a mathematical framework for understanding the asymmetry between matter and antimatter in the Universe. This framework is governed by an equation that describes the baryon-to-entropy ratio as follows:
\begin{equation}\label{35}
\frac{\eta_{B}}{s}\simeq -\bigg(\frac{15g_{b}}{4\pi^{2}g_{s}}\bigg)\bigg[\frac{\dot{R}}{M_{\star}^{2}T_{D}}\bigg],
\end{equation}
The variable $g_{s}$ in this equation stands for the total degrees of freedom of massless particles and $g_{b}$ represents the intrinsic degrees of freedom that are unique to baryons. The symbol $\dot{R}$ represents the rate of change of the Ricci scalar over time, while $M_{\star}$ denotes the energy threshold at which processes that violate CP symmetry emerge. The equation suggests that the baryon-to-entropy ratio is sensitive to the dynamics of the Ricci scalar, establishing a fundamental connection between the cosmos's expanding nature and the generation of baryon asymmetry. During the radiation-dominated era, the energy density and temperature $T$ are connected through a fundamental statistical relation, enabling a clear understanding of their interdependence as follows:
\begin{equation}\label{36}
\rho(T)=\frac{\pi^{2}}{30}g_{s}T^{4}.
\end{equation}
This equation reveals the temperature-dependent scaling behavior of energy density in the primordial Universe, providing insight into the evolutionary dynamics of the cosmos. Following this, we incorporate an interaction term that breaks CP symmetry, giving rise to baryon asymmetry in the context of $f(Q,L_{m})$ gravity. The interaction term is mathematically represented as:
\begin{equation}\label{37}
\frac{1}{M_{\star}^{2}}\int \sqrt{-g}d^{4}x(\partial_{\mu}(Q+L_{m}))J^{\mu},
\end{equation}
where, $J^{\mu}$ signifying the baryonic current that characterizes the dynamics of baryon number. This term triggers the violation of CP symmetry, which in turn induces the necessary conditions for the creation of a baryon-antibaryon asymmetry in the early Universe, paving the way for the observed imbalance. The baryon-to-entropy ratio in $f(Q,L_{m})$ gravity is governed by the following equation, which connects the baryon asymmetry to the $Q$ scalar and matter fields:
\begin{equation}\label{38}
\frac{\eta_{B}}{s}\simeq -\bigg(\frac{15g_{b}}{4\pi^{2}g_{s}}\bigg)\bigg[\frac{\dot{Q}+\dot{L}_{m}}{M_{\star}^{2}T_{D}}\bigg]
\end{equation}
The equation (\ref{38}) states that the baryon-to-entropy ratio is proportional to the ratio of the time derivatives of $Q$ and $L_{m}$, scaled by the energy scale $M_{\star}$ and temperature $T_{D}$. The term $(\frac{15g_{b}}{4\pi^{2}g_{s}})$ is a numerical factor that depends on the degrees of freedom for baryons and entropy and the negative sign indicates that the baryon-to-entropy ratio is generated by the violation of CP symmetry (which is encoded in the $Q$ and $L_{m}$ terms).

By inserting the expression for the Hubble parameter $H(t)$ from equation (\ref{20}) into the field equation (\ref{17}), we arrive at the following expression for energy density:
\begin{equation}\label{39}
\rho=\frac{\alpha}{\beta}6^{n}(1-2n)\bigg(\frac{2n}{3(t-t_{0})}\bigg)^{2n},
\end{equation}
By equating the right-hand sides of equations (\ref{39}) and (\ref{36}), we can solve for the decoupling time $t_{D}$ in terms of the temperature $T_{D}$, obtaining:
\begin{equation}\label{40}
t_{D}=t_{0}+\bigg[\frac{30\alpha2^{3n}(1-2n)n^{2n}}{\pi^{2}g_{s}T_{D}^{4}\beta3^{n}}\bigg]^{\frac{1}{2n}},
\end{equation}
Equation (\ref{38}) is utilized to compute the baryon-to-entropy ratio for our model, resulting in:
\begin{equation}\label{41}
\frac{\eta_{B}}{s}\simeq -\frac{15g_b}{4\pi^{2}g_{\star}M_{\star}^{2}T_{D}}\bigg[-16n^{2}\bigg(\frac{\pi^{2}g_{\star}T_{D}^{4}\beta3^{n}}{30\alpha2^{3n}(1-2n)n^{2n}}\bigg)^{\frac{3}{2n}}-\frac{\alpha}{\beta}\bigg(\frac{2}{3}\bigg)^{n}
(1-2n)(2n)^{2n+1}\bigg(\frac{\pi^{2}g_{\star}T_{D}^{4}\beta3^{n}}{30\alpha2^{3n}(1-2n)n^{2n}}\bigg)^{\frac{2n+1}{2n}}\bigg].
\end{equation}
\begin{figure}[h!]
\centering
  \includegraphics[scale=0.5]{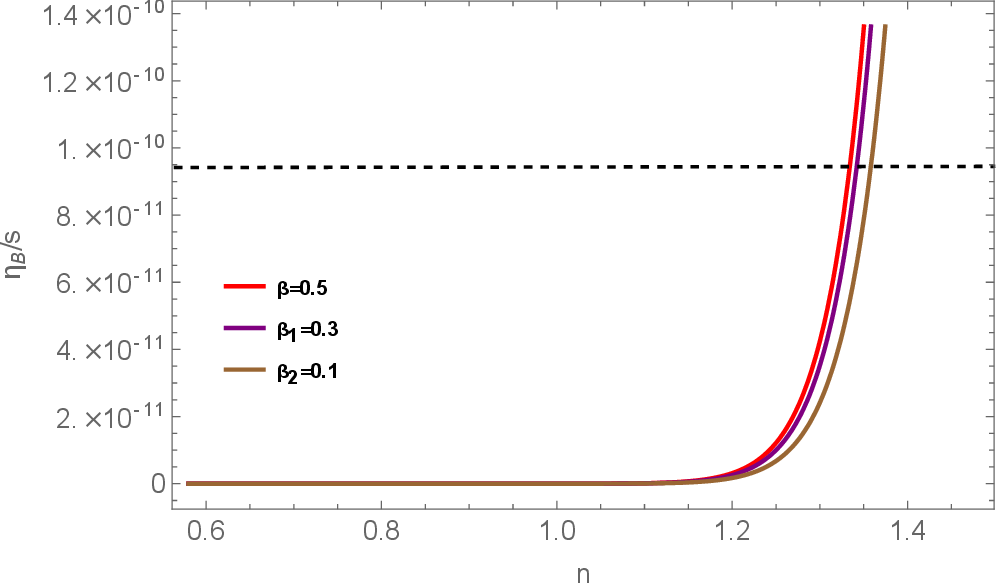}
  \caption{The plot displays $\frac{\eta_{B}}{s}$ as a function of $n$ for our model, with varying values of $\beta$ and fixed parameters: $g_{b}=1$, $g_{s}=106$, $T_{D}=2\times 10^{16}\; GeV$, $M_{\star}=2\times 10^{12}\; GeV$ and $\alpha\simeq -1.95084\times10^{86}$.}
  \label{fig:f6}
\end{figure}\\
\begin{table}[h!]
\centering
\caption{The values of $\frac{\eta_{B}}{s}$ for different values of the model parameters}
\begin{tabular}{||p{1.7cm}|p{2.0cm}|p{2.8cm}||}
\hline\hline
\hspace{0.7cm}$\beta$ & \hspace{0.7cm}$n$ & \hspace{1.2cm}$\frac{\eta_{B}}{s}$ \\[1pt]
\hline\hline
\hspace{0.6cm}$0.5$ & \hspace{0.5cm}$1.33465$ & \hspace{0.3cm}$9.422\times 10^{-11}$\\[1pt]
\hline
\hspace{0.6cm}$0.3$ &  \hspace{0.5cm}$1.33986$ & \hspace{0.3cm}$9.4173\times 10^{-11}$\\[1pt]
\hline
\hspace{0.6cm}$0.1$ & \hspace{0.5cm}$1.35787$ & \hspace{0.3cm}$9.42023\times 10^{-11}$\\[1pt]
\hline\hline
\end{tabular}
\label{Tab:T4}
\end{table}

It is crucial to recognize that $n$ cannot be equal to $\frac{1}{2}$, as this would imply a vanishing baryon-to-entropy ratio ($\frac{\eta_{B}}{s}$), which is physically implausible. Through a precision calculation using equation (\ref{41}), we determine the value of $\alpha$ that reproduces the observed baryon-to-entropy ratio. By inputting the model parameters $g_{b}=1$, $g_{s}=106$, $T_{D}=2\times 10^{16}\; GeV$, $M_{\star}=2\times 10^{12}\; GeV$, $\beta=0.5$ and $n=0.63$, we find $\alpha\simeq -1.95084\times10^{86}$. This result is in excellent agreement with the observed value of $\frac{\eta_{B}}{s}\simeq 9.42\times10^{-11}$, showcasing the model's predictive power and providing a strong constraint on the parameter $\alpha$. The graphical analysis in Figure \ref{fig:f6} illustrates the sensitivity of the $\frac{\eta_{B}}{s}$ to variations in the model parameters $\beta$ and $n$. The convergence point of the curves in the figure, denoted by the dotted line, represents the benchmark value of the baryon-to-entropy ratio, $\frac{\eta_{B}}{s}\simeq 9.42\times10^{-11}$ \cite{DN07}. The table \ref{Tab:T4} provides a concise summary of the model's predictions for $\frac{\eta_{B}}{s}$, systematically exploring the impact of different $n$ and $\beta$ values on the baryon-to-entropy ratio and facilitating a comprehensive evaluation of the model's performance. The parameter $n$ and $\beta$ exhibit a reciprocal relationship, where an increment in $n$ necessitates a corresponding decrement in $\beta$, thereby facilitating the attainment of congruent values for $\frac{\eta_{B}}{s}$. The plot reveals values of $n$ as $1.33465, 1.33986$ and $1.35787$, which are in close agreement with the values of $n$ obtained from observations, particularly for the phantom dataset where $n=1.46^{+0.03}_{-0.03}$. This consistency indicates that the model parameters used in our analysis successfully reproduce the trends observed in the phantom dataset.
\section{Generalized Baryogenesis in modified $f(Q,L_{m})$ gravity}\label{sec6}
\hspace{0.6cm} In the context of generalized gravitational baryogenesis for $f(Q,L_{m})$ gravity, the CP-violating interaction can be formulated by incorporating both the non-metricity scalar $Q$ and matter-Lagrangian $L_{m}$. The interaction term responsible for generating the baryon asymmetry is then given by:
\begin{equation}\label{42}
\frac{1}{M_{\star}^{2}}\int \sqrt{-g}J^{\mu}d^{4}x(\partial_{\mu}(Q+L_{m})).
\end{equation}
In this framework, the baryon-to-entropy ratio is expressed as:
\begin{equation}\label{43}
\frac{\eta_{B}}{s}\simeq -\frac{15g_{b}}{4\pi^{2}g_{s}}\bigg(\frac{\dot{Q}f_{Q}+\dot{L}_{m}f_{L_{m}}}{M_{\star}^{2}T_{D}}\bigg).
\end{equation}
where $f_{Q}=\frac{df}{dQ}$ and $f_{L_{m}}=\frac{df}{dL_{m}}$. This formula describes how the baryon asymmetry evolves as a result of the time dependence of $Q$ and $L_{m}$, captured by their time derivatives $\dot{Q}$ and $\dot{L}_{m}$. The non-metricity scalar's interaction with the matter Lagrangian triggers a gravitational baryogenesis effect, which effectively creates the observed dominance of matter over antimatter in the Universe. The expression of $\frac{\eta_{B}}{s}$ for generalized gravitational baryogenesis is derived from the equations (\ref{43}) and (\ref{20}) as follows:
\begin{equation}\label{44}
\frac{\eta_{B}}{s}\simeq -\frac{15g_{b}}{4\pi^{2}g_{s}}\bigg[-\frac{n^{3}\alpha2^{3n+1}}{3^{n-1}}\bigg(\frac{\pi^{2}g_{\star}T_{D}^{4}\beta3^{n}}{30\alpha2^{3n}(1-2n)n^{2n}}\bigg)^{\frac{2n+1}{2n}}-\alpha\bigg(\frac{2}{3}\bigg)^{n}
(1-2n)(2n)^{2n+1}\bigg(\frac{\pi^{2}g_{\star}T_{D}^{4}\beta3^{n}}{30\alpha2^{3n}(1-2n)n^{2n}}\bigg)^{\frac{2n+1}{2n}}\bigg].
\end{equation}
\begin{table}[h!]
\centering
\caption{The values of $\frac{\eta_{B}}{s}$ for different values of the model parameters}
\begin{tabular}{||p{1.7cm}|p{2.0cm}|p{2.8cm}||}
\hline\hline
\hspace{0.7cm}$\beta$ & \hspace{0.7cm}$n$ & \hspace{1.2cm}$\frac{\eta_{B}}{s}$ \\[1pt]
\hline\hline
\hspace{0.6cm}$0.5$ & \hspace{0.5cm}$1.33398$ & \hspace{0.3cm}$9.42054\times 10^{-11}$\\[1pt]
\hline
\hspace{0.6cm}$0.3$ &  \hspace{0.5cm}$1.33716$ & \hspace{0.3cm}$9.42023\times 10^{-11}$\\[1pt]
\hline
\hspace{0.6cm}$0.1$ & \hspace{0.5cm}$1.35399$ & \hspace{0.3cm}$9.41229\times 10^{-11}$\\[1pt]
\hline\hline
\end{tabular}
\label{Tab:T5}
\end{table}
\begin{figure}[h!]
\centering
  \includegraphics[scale=0.5]{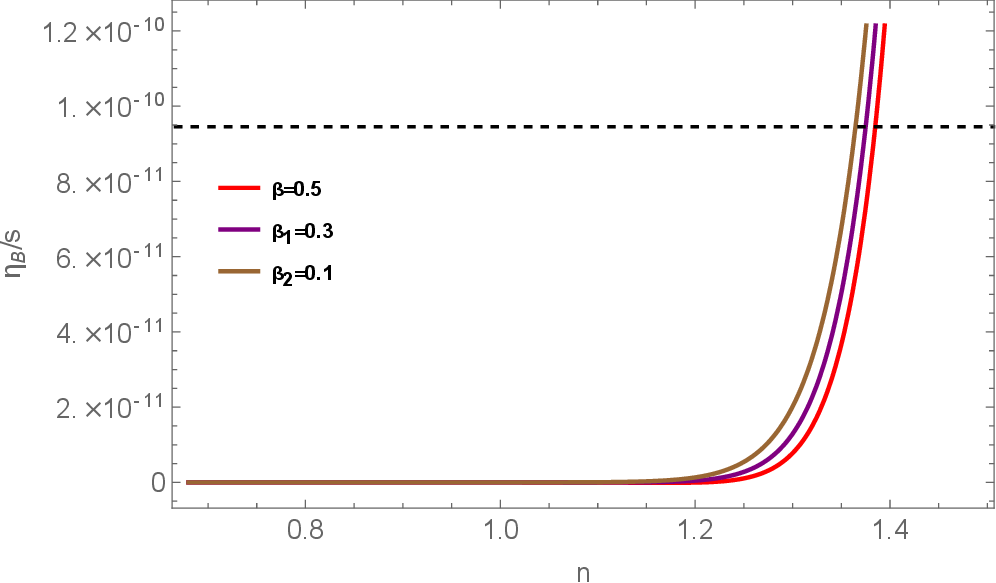}
  \caption{The plot displays $\frac{\eta_{B}}{s}$ as a function of $n$ for our model, with varying values of $\beta$ and fixed parameters: $g_{b}=1$, $g_{s}=106$, $T_{D}=2\times 10^{16}\; GeV$, $M_{\star}=2\times 10^{12}\; GeV$ and $\alpha\simeq -1.95084\times10^{86}$.}
  \label{fig:f7}
\end{figure}

It's essential to understand that setting $n=\frac{1}{2}$ would lead to a zero baryon-to-entropy ratio $\frac{\eta_{B}}{s}$, which is not physically viable. Figure \ref{fig:f7} illustrates the variation of the baryon-to-entropy ratio as a function of $n$, for three different values of $\beta$, showcasing a intriguing dependence of the $\frac{\eta_{B}}{s}$ on both $n$ and $\beta$. The intersection of the $\frac{\eta_{B}}{s}$ curves with the dashed line, which reflects the observed baryon-to-entropy ratio, occurs within the range $n=1.32965$ to $n=1.39252$. This suggests that the model parameters predict a value of $\frac{\eta_{B}}{s}$ that is both positive and consistent with the observational upper limit of $9.42\times10^{-11}$, given $\alpha\simeq -1.95084\times10^{86}$. The alignment of the predicted results with the observational constraints reinforces the reliability of the model in capturing the physics of baryogenesis.  

The table \ref{Tab:T5} offers a detailed breakdown of how varying $n$ and $\beta$ impact the baryon-to-entropy ratio $\frac{\eta_{B}}{s}$, highlighting the model's predictive accuracy. The relationship between these parameters is such that as $n$ increases, $\beta$ must be decrease to achieve similar values for $\frac{\eta_{B}}{s}$. The calculated values of $n (1.33465, 1.33986\; \& \; 1.35787)$ fall in close proximity to the observational value for the phantom dataset, $n=1.46^{+0.03}_{-0.03}$. This alignment suggests that the model is adept at replicating the baryogenesis behavior seen in the data, affirming its robustness and capacity to mirror real-world cosmological trends.
\section{Conclusion}\label{sec7}
\hspace{0.6cm} In this paper, we have proposed and analyzed a cosmological model based on $f(Q,L_{m})$ gravity, where the functional form is taken as $f(Q,L_{m})=\alpha Q^{n}+\beta L_{m}$. By using various observational datasets, including the Hubble $46$, BAO $15$ and Phantom datasets, we have constrained the model parameters and explored its implications on the cosmic expansion history and baryogenesis. The analysis led to the following key parameter estimates: For the Hubble dataset: $H_{0}= 67.17^{+0.83}_{-0.82}$ $km/s/Mpc$ and $n=0.68^{+0.02}_{-0.02}$, $H_{0}= 68.19^{+0.81}_{-0.82}$ $km/s/Mpc$ and $n=0.68^{+0.01}_{-0.02}$ for the Hubble+BAO datasets and $H_{0}= 69.22^{+0.83}_{-0.83}$ $km/s/Mpc$ and $n=1.46^{+0.03}_{-0.03}$ for the Hubble+Phantom datasets. These values align with current observations and provide a consistent framework for studying the Universe's expansion rate, as well as the role of non-metricity in modifying gravity.

Additionally, we extended the analysis to investigate baryogenesis within the $f(Q,L_{m})$ gravity model. Utilizing the constrained parameters, we investigated the baryon-to-entropy ratio $\frac{\eta_{B}}{s}$ in the context of $f(Q,L_{m})$ gravity, examining how this modified gravity theory can provide a viable explanation for the observed baryon asymmetry in the Universe. Our results demonstrate that the model successfully reproduces the observed value of $\frac{\eta_{B}}{s}\simeq9.42\times10^{-11}$ for particular values of the parameters $\beta$, $n$ and $\alpha\simeq -1.95084\times10^{86}$, especially when considering values of $n$ close to the radiation-dominated phase of the Universe $(n\neq\frac{1}{2})$. Moreover, the figure reveals that when the parameter $n$ assumes the values $1.33465, 1.33986$ and $1.35787$, which are remarkably close to the observational limits, the model's predictions align with the observed baryon asymmetry.   

In addition to the standard gravitational baryogenesis, we also explored the generalized gravitational baryogenesis scenario within the context of $f(Q,L_{m})$ gravity. From our analysis, we found that the predicted values of $\frac{\eta_{B}}{s}$  align well with the observational constraints, particularly in the range $1.32965<n<1.39252$, where the calculated ratio intersects the observed value of $9.42\times10^{-11}$. The relationship between $n$ and $\beta$ reveals that as $n$ increases, $\beta$ must be decrease to maintain agreement with the observed $\frac{\eta_{B}}{s}$, underscoring the delicate balance between these parameters. Moreover, the values of $n$ derived from our model $1.33465,1.33986$ and $1.35787$ closely match the observational value of $n=1.46^{+0.03}_{-0.03}$ obtained from the phantom dataset. This close agreement between theoretical predictions and observational data highlights the model's ability to accurately describe the baryogenesis process and reinforce its reliability in capturing key cosmological phenomena.

\end{document}